\documentstyle[12pt,aasms4]{article}

\input{epsfig.sty}

\newcommand{\be}{\begin{equation}}
\newcommand{\ee}{\end{equation}}

\lefthead{Huang }
\righthead{Afterglows from Soft Gamma Repeaters}

\begin{document}


\title{TESTING THE FIREBALL/BLASTWAVE MODEL \\ 
       BY MONITORING AFTERGLOWS FROM SOFT GAMMA \\ 
       REPEATERS \footnote{ Received: 1998 \hspace{2mm} --- \hspace{2mm} ---  (A detailed version of astro-ph/9810449)\\ 
       {\large\bf Published in ``Journal of Nanjing University'', 1998, Vol. 34, 96} }
       }
\author{{\it Y. F. Huang}}
\affil{ ( Department of Astronomy, Nanjing University, Nanjing 210093, 
          P.R. China ) }
\authoraddr{ ( Department of Astronomy, Nanjing University, Nanjing 210093, 
	       P.R. China ) }

\baselineskip=7mm
\noindent 
{\bf Abstract} \hspace{3mm}
The popular fireball/blastwave model of classical $\gamma$-ray bursts
is applied to soft $\gamma$-ray bursts. It is found that
X-ray afterglows from strong events may be above their quiescent levels for
40 $-$ 400 seconds. Optical afterglows may also be detectable.
By monitoring the three repeaters, we will have an ideal way to check
the fireball/blastwave model. 

\noindent
{\bf Keywords} \hspace{3mm} gamma rays: bursts,  shock waves 

\noindent 
{\bf Classification Index} \hspace{3mm} P142.6, P144.6

\vspace{8mm}

\noindent
{\large \bf 0 \hspace{3mm} INTRODUCTION}

\vspace{4mm}

The recent detection of afterglows from some $\gamma$-ray bursts (GRBs) 
located by BeppoSAX opens up a new era in the studies of GRBs.$^{[1-3]}$ 
Afterglows were detected in X-rays from GRB 970228, 970402, 970508, 970616, 
970828, in optical band from GRB 970228, 970508, and even in radio from 
GRB 970508.The possible host galaxy of GRB 970228 and the 
determined redshift $0.835 < z < 2.1$ for GRB 970508  
strongly indicate a cosmological origin for GRBs. 

GRBs might be produced by highly relativistic fireballs.$^{[4,5]}$ 
After the main GRB, the collision between the GRB ejecta and the
interstellar medium (ISM) provides a natural explanation for the 
power-law decay of the observed low energy afterglows.$^{[6-10]}$
However, so few GRBs have
been located rapidly and accurately enough for us to search for their
afterglows, that the cosmological origin of GRBs and the correctness
of the fireball/blastwave model still need more tests. GRBs 
occuring at a definite distance and in a fixed direction would be ideal
for checking the model. Luckily enough, we do have such ideal
objects at hand.

While the nature of the so called ``classical $\gamma$-ray bursts'' is
still controversial, cases for a subtle class of GRBs, the soft 
$\gamma$-ray repeaters (SGRs), are much clearer. SGRs are characterized 
mainly by their soft spectrums and unpredicatable recurrences.$^{[11]}$  
There are only three known SGRs (0526$-$66, 1806$-$20 and 1900$+$14), 
all have been tentatively associated with supernova remnants (SNRs).
Recently a possible fourth SGR was reported, but need to be confirmed.
A typical SGR burst lasts several hundred milliseconds, emitting 
$\sim 10^{40} - 10^{41}$ ergs in soft $\gamma$-rays. Due to
the huge energy, the limited volume and the small timescale, a 
fireball seems inevitable before soft $\gamma$-rays are emitted, just
as a cosmological GRB. This has led to our suggestion that we could
check the fireball model by monitoring the SGR sources. Below  
the fireball/blastwave model is first briefly described and then applied  
to SGR bursts to predict their afterglows in X-ray and optical bands.

\vspace{8mm}

\noindent 
{\large \bf 1 \hspace{3mm} AFTERGLOWS FROM SGR BURSTS}

\vspace{4mm}

\noindent 
{\bf 1.1  the Adiabatic Expansion}

\vspace{2mm}

A fireball with total initial energy $E_0$ and initial bulk Lorentz 
factor $\eta \equiv E_0/M_0 c^2$, where $M_0$ is the initial baryon 
mass, is expected to radiate half of its energy 
in $\gamma$-rays during the GRB phase, either due 
to an internal-shock or an external-shock mechanism. 
The subsequent expansion generates an ultrarelativistic shock. The 
Lorentz factor of the shock ($\Gamma$) and the shocked ISM ($\gamma$) are
related by $\Gamma^2 \approx 2 \gamma^2$. In the shell's comoving frame, 
number density ($n'$) and energy density ($e'$) of the shocked ISM are 
$n' \approx 4 \gamma n$ and $ e' \approx 4 \gamma^2 n m_{\rm p} c^2$, 
respectively,
where $n$ is the number density of the unshocked ISM.$^{[12]}$ 
In the case of adiabatic expansion, energy is conserved, then we
get a useful expression for $\gamma$ and $R$ (shock radius), 
\begin{equation}
\gamma^2 R^3 \approx E_0/(8 \pi n m_{\rm p} c^2).
\end{equation}

Photons observed within a time interval of $dt$ are in fact emitted within an
interval of $dt_{\rm b} = dt / (1 - v/c) \approx 2 \gamma^2 dt$ in the
burster's fixed frame, where $v$ is the observed velocity of the shocked
ISM. Then $R$ and $t$ are related by 
\begin{equation}
\frac{dR}{dt} \approx 2 \gamma^2 c.
\end{equation}
Under the assumption that $\gamma \gg 1$, combining equation (1) and 
equation (2), we can derive a simple solution:
\begin{equation}
R(t) \approx 8.93 \times 10^{15} E_{51}^{1/4} n_1^{-1/4} t^{1/4} {\rm cm} 
 = 5.02 \times 10^{13} E_{42}^{1/4} n_1^{-1/4} t^{1/4} {\rm cm},
\end{equation}
\begin{equation}
\gamma (t) \approx 193 E_{51}^{1/8} n_1^{-1/8} t^{-3/8} 
 = 14.5 E_{42}^{1/8} n_1^{-1/8} t^{-3/8},
\end{equation}
where $E_0 = 10^{51} E_{51}$ ergs $= 10^{42} E_{42}$ ergs, $n = n_1$    
cm$^{-3}$ and $t$ is in unit of second. 
For a more accurate solution please see Huang {\em et al.}'s numerical 
evaluation.$^{[10]}$

\vspace{5mm}

\noindent
{\bf 1.2  Synchrotron Radiation}

\vspace{2mm}

Electrons in the shocked ISM are highly relativistic. Inverse Compton 
cooling of the electrons may not contribute to emission in X-ray and 
optical bands we are interested in. Only synchrotron radiation will be
considered below. The electron distribution in the shocked ISM is assumed
to be a power-law function of electron energy, as expected for shock
acceleration, 
\begin{equation}
dn_{\rm e}' / d \gamma_{\rm e} \propto \gamma_{\rm e}^{-p}, 
\gamma_{\rm e,min} \leq \gamma_{\rm e} \leq \gamma_{\rm e,max}, 
\end{equation}
where $\gamma_{\rm e,min}$ and $\gamma_{\rm e,max}$ are the minimum and
maximum Lorentz factors of electrons, and $p$
is an index varying between 2 and 3. I suppose that the magnetic field
energy density (in the comoving frame) is a fraction $\xi_{\rm B}^2$ of the
energy density, $B'^{2}/8 \pi = \xi_{\rm B}^{2} e'$, and that the electron 
carries a fraction $\xi_{\rm e}$ of the energy, 
$\gamma_{\rm e,min} = (m_{\rm p}/m_{\rm e}) \xi_{\rm e} \gamma + 1$.

The spectral property of synchrotron radiation from such a collection of
electrons is clear. In the comoving frame, the
characteristic photon frequency 
is $\nu_{\rm m} = e B' \gamma_{\rm e,min}^2 / (2 \pi m_{\rm e} c)$,
where $e$ is the electron charge. The spectral peaks at
$\nu_{\rm max} \approx 0.29 \nu_{\rm m}$. 
For frequency $\nu \gg \nu_{\rm max}$, and
$\nu \ll \nu_{\rm max}$, the flux density scales as $\nu^{- \alpha}$ and
$\nu^{1/3}$ respectively, where $\alpha = (p-1)/2$. 

The specific intensity at frequency $\nu$ in the comoving frame
($I_{\nu,{\rm co}}$) can be transformed into 
the observer's frame by the following
equations:
\begin{equation}
\nu_{\oplus} = (1 + v/c) \gamma \nu,
\end{equation}
\begin{equation}
I_{\nu_{\oplus},\oplus} = (1 + v/c) \gamma^3 I_{\nu,{\rm co}}.
\end{equation}
Then the observed flux density is 
$S_{\nu_{\oplus},\oplus} = \pi (k \gamma c t)^2
I_{\nu_{\oplus},\oplus}/D^2$, where $k \approx 3$ is a coefficient 
introduced to correct for the effect on the observed emitting surface by
the dynamical deceleration.$^{[10]}$
The flux observed by a detector is an integral of 
$S_{\nu_{\oplus},\oplus}$ over the range between lower and 
upper frequency limits of the detector.

\vspace{5mm}

\noindent 
{\bf 1.3 Numerical Results}

\vspace{2mm}

I have carried out detailed numerical evaluation to 
investigate the afterglows from SGR bursts, following 
Huang {\em et al.}'s simple model.$^{[10]}$ I chose $E_0$ 
between $10^{40}$ ergs and $10^{42}$ ergs, and $n=1$ or 10 cm$^{-3}$.
In each case I set $p=2.5$, $\xi_{\rm e} = 0.1$ and $d = 10$ kpc. 
Since $M_0$ is 
a parameter having little influence on the afterglows,  
I chose $M_0$ so that $\eta \approx 280$ in all cases. X-ray flux 
($F_{\rm X}$) is integrated from 0.1 keV to 10 keV, and optical flux 
densities for R band ($S_{\rm R}$) are calculated. 
The evolution of $F_{\rm X}$ and $S_{\rm R}$ are illustrated 
in Figures 1 and 2 respectively. We see that for a strong 
burst ($E_0 > 10^{41}$ ergs), $F_{\rm X}$ can in general keep to be
above $10^{-12}$ ergs$\cdot$cm$^{-2}\cdot$s$^{-1}$ for 40 $-$ 200 seconds
and $S_{\rm R}$ can keep to be 
above $10^{-29}$ ergs$\cdot$cm$^{-2}\cdot$s$^{-1}\cdot$Hz$^{-1}$ 
(corresponding to $R \approx 24^{\rm m}.0$) for 200 $-$ 1000 seconds. 
But if we take
$E_0 = 10^{40}$ ergs, then $F_{\rm X}$ can hardly be greater than 
$2 \times 10^{-12}$ ergs$\cdot$cm$^{-2}\cdot$s$^{-1}$.

To compare the afterglows from cosmological, Galactic Halo and 
SGR bursts more directly, we plot their X-ray and optical light
curves in contrast in Figures 3 and 4. It is clearly shown that
the predicted afterglows from cosmological GRBs last much longer than
any kind of Galactic bursts. This is consistent with previous 
conclusion. Since X-ray afterglows are observed more than a week later
and optical afterglow is observed even more than six months later for
GRB 970228, we suggest that the observed time scales of afterglows is
another strong evidence favoring the cosmological origin.

\vspace{8mm}

\noindent
{\large \bf 2 \hspace{3mm} COMPARISION BETWEEN PREDICTION AND OBSERVATION}

\vspace{4mm}

The three known SGRs have been extensively looked after in X-ray, optical
and radio bands. A pointlike X-ray source has been identified 
associating with each SGR, but only SGR 1806$-$20 has a 
detectable optical counterpart. Below is a brief review.$^{[13]}$ 

SGR 0526$-$66 is associated with SNR N49, about 55 kpc from us.
A permanent X-ray hot spot is found with an unabsorbed flux of 
$\sim 2.0 \times 10^{-12}$ ergs$\cdot$cm$^{-2}\cdot$s$^{-1}$ 
(0.1 $-$ 2.4 keV). 
No optical counterpart has been identified.
Dickel et al. placed $3 \sigma$ upper limits on the radio
(less than 0.3 Jy at 12.6 cm), infrared (less than 39 $\mu$Jy and 58
$\mu$Jy at 2.16 and 1.64 $\mu$m respectively), and optical (less than
40 $\mu$Jy at 656.3 nm) emission from the X-ray source.

SGR 1806$-$20 is associated with the Galactic SNR G10.0$-$0.3, 10
to 15 kpc from the Earth. A steady pointlike X-ray source with an unabsorbed 
flux of $\sim 10 \times 10^{-12}$ ergs$\cdot$cm$^{-2}\cdot$s$^{-1}$
has been observed.  Optical observations have revealed a 
luminous O/B type companion to this SGR.  However, 
due to a giant molecular cloud located at this direction, interstellar
extinction is serious (A$_{\rm v}$ = 30$^{\rm m}$), 
and the optical source is heavily reddened.

The least active source SGR 1900$+$14 is associated with the Galactic 
SNR G42.8$+$0.6, 7 to 14 kpc from us. A quiescent, steady, point 
X-ray source is present at its position, with an unabsorbed flux 
of $3.0 \times 10^{-12}$ ergs$\cdot$cm$^{-2}\cdot$s$^{-1}$. 
No optical source is detected down to limiting magnitude of 
$m_{\rm v} \approx 24^{\rm m}.5$. 

In order to be detectable, the X-ray afterglow flux from a SGR burst 
should at least be comparable to that of the quiescent X-ray source.
Taken $10^{-12}$ ergs$\cdot$cm$^{-2}\cdot$s$^{-1}$ as a threshold, then the
predicted afterglows will generally be above the value for about
40 $-$ 200 seconds for intense events (Figure 1). 
Since the peak flux can be as high
as $10^{-8} - 10^{-7}$ ergs$\cdot$cm$^{-2}\cdot$s$^{-1}$, such an afterglow
should be observable by those satellites now in operation, such as 
ROSAT and ASCA. If detected, afterglows from SGR bursts would be ideal
to test the fireball/blastwave model. We suggest that SGRs should be
monitored during their active periods. Cases are similar for optical
afterglows. If we 
took $S_{\rm R} = 100$ $\mu$Jy ($m_{\rm R} \approx 19^{\rm m}$) as the
threshold, afterglow would last less than 100 seconds, but if we took
$S_{\rm R} = 1$ $\mu$Jy ($m_{\rm R} \approx 24^{\rm m}$), 
then we would have several $10^3$ seconds (Figure 2).

We notice that some researchers do have monitored the SGRs in optical
and radio wavelengths. After monitoring SGR 1806$-$20 with the VLA 
in 10 epochs spreading over 5 months, Vasisht {\em et al.} reported 
that there was no radio variability 
above the 25\% level on postburst timescales ranging from 
2 days to 3 months. Radio afterglows are beyond our discussion here 
because strong self-absorption is involved. 
Pedersen {\em et al.} have reported three possible optical flashes 
from SGR 0526$-$66, but none of their light curves shows any sign of 
afterglows. We think it was either due to the limited aperture (50 cm)
of their telescope or that maybe the flashes were spurious. The latter
seems more possible since no soft $\gamma$-ray bursts were observed 
simultaneously.

Of special interest is the most prolific source SGR 1806$-$20. On 
1993 October 9.952414 UT a soft $\gamma$-ray burst occured.  ASCA 
satellite happened to be observing the SGR at that moment and 
recorded a simultaneous X-ray burst.
Sonobe {\em et al.} pointed out that there were
no obvious mean intensity changes in X-rays prior to the burst nor 
following the burst not only on a timescale of 1 day, but also on 
timescales of minutes.$^{[14]}$ This is not inconsistent with 
our predictions since it was a relatively weak burst, 
with $E_0$ about $10^{39}$ ergs. Afterglows from this burst are 
not expected to be detectable.

We have also calculated afterglows from such a unique burst as 
GRB 790305 from SGR 0526$-$66,$^{[15]}$
taking $E_0 = 10^{45}$ ergs and $d = 55$ kpc.
The light curves are plotted in Figures 5 and 6. It is found 
that the X-ray afterglows should be detectable 
($> 10^{-12}$ ergs$\cdot$cm$^{-2}\cdot$s$^{-1}$) for 
several hours, and $S_{\rm R}$ will be 
above 100 $\mu$Jy ($m_{\rm R} \approx 19^{\rm m}$) for about one hour.
Had the source been monitored on 1979 March 5, afterglows should 
have been observed.

\vspace{8mm}

\noindent 
{\large \bf 3 \hspace{3mm} DISCUSSION AND CONCLUSIONS}

\vspace{4mm}

Gamma-ray bursts occuring at three different distance scales have
been observed or suggested: Classical GRBs at cosmological distances,
Classical GRBs in the Galactic Halo,
and SGRs at about 10 kpc distances.  The cosmological
origin of Classical GRBs and the fireball/blastwave model are
two propositions. Although they are consistent with each other in 
that the observed power-law decays of afterglows from GRBs can be
naturally explained,  
both of them are in urgent need of more observational
tests, especially independent ones. The possible host galaxy of 
GRB 970228 and the red shift of 
GRB 970508 are two strong proofs for
the cosmological origin, but they are far from enough. Here we have 
stressed that the observed afterglow timescale (more than one week
in X-rays and six months in optical band) is another strong proof, 
since afterglows from any kind of Galactic GRBs will be too weak to
be viable on that timescale, as clearly shown in this paper.

Soft $\gamma$-ray bursts from SGRs might be good 
candidates to be used to test the fireball/blastwave model independently.
The arguments are obvious: the distances are much certain, their
accurate positions are available, they burst out repeatedly, their
origins are relatively clear so that we feel more confident about 
them. According to our calculations, afterglows from a strong 
SGR burst will generally be detectable. It is thus suggested that
the SGRs should be monitored during their active periods. Although 
such observations are imaginably difficult, the results will 
be valuable, not only in that the afterglows might be acquired and 
the fireball/blastwave model could be tested, but also that the 
simultaneous bursting behaviors in X-ray and optical wavelengths other
than soft $\gamma$-rays are important to our understanding of the
SGRs themselves.

We particularly noticed an X-ray burst from SGR 1806$-$20 detected by
the ASCA satellite. It is a great pity that the corresponding soft
$\gamma$-ray burst is rather weak so that no afterglow was observed.
However, the negative detection of X-ray afterglows itself may
be regarded as a proof supporting the fireball/blastwave model, 
although it is a relatively weak one.

\vspace{10mm}

\newpage

\begin{center}
{\large \bf References}
\end{center}

\begin{description}
\baselineskip=2mm
\item {[1] Fishman G J, Meegan C A. Gamma-ray bursts. Annu Rev Astron
	   Astrophys, 1995, 33: 415 }
\item {[2] van Paradijs J, Groot P J, Galama T, {\em et al.} 
	   Transient optical emission from the error box of the 
	   $\gamma$-ray burst of 28 February 1997. Nature, 1997, 386: 686 }
\item {[3] Djorgovski S G, Metzger M R, Kulkarni S R, {\em et al.} 
	   The optical counterpart to the $\gamma$-ray burst 
	   GRB 970508. Nature, 1997, 387: 876 }
\item {[4] M\'{e}sz\'{a}ros P, Rees M J. Gamma-ray bursts: multiwaveband 
	   spectral predictions for blastwave models. Astrophys J, 1993, 
	   418: L59 }
\item {[5] Paczy\'{n}ski B, Rhoads J E. Radio transients from Gamma-ray 
	   bursters. Astrophys J, 1993, 418: L5 }
\item {[6] M\'{e}sz\'{a}ros P, Rees M J. Optical and long wavelength 
	   afterglow from gamma-ray bursts. Astrophys J, 1997, 476: 232 }
\item {[7] Wijers R A M J, Rees M J, M\'{e}sz\'{a}ros P. Shocked by 
	   GRB 970228: the afterglow of a cosmological fireball. Mon Not 
	   R Astron Soc, 1997, 288: L51 }
\item {[8] Tavani M. X-ray afterglows from Gamma-ray bursts. 
	   Astrophys J, 1997, 483: L87 }
\item {[9] Waxman E. $\gamma$-ray burst afterglow: confirming the
	   cosmological fireball model. Astrophys J, 1997, 489: L33 }
\item {[10] Huang Y F, Dai Z G, Wei D M, Lu T. Gamma-ray bursts: postburst 
	    evolution of fireballs. Mon Not R Astron Soc, 1998, in press }
\item {[11] Norris J P, Hertz P, Wood K S, Kouveliotou C. 
            On the nature of soft gamma repeaters. 
            Astrophys J, 1991, 366: 240 }
\item {[12] Blandford R D, McKee C F. Fluid dynamics of relativistic 
	    blast waves. The Physics of Fluids, 1976, 19: 1130 }
\item {[13] Duncan R C, Li H. The halo beaming model for gamma-ray 
	    bursts. Astrophys J, 1997, 484: 720 }
\item {[14] Sonobe T, Murakami T, Kulkarni S R, Aoki T, Yoshida A. 
	    Characteristics of the persistent emission of 
	    SGR 1806$-$20. Astrophys J, 1994, 436: L23 }
\item {[15] Cline T L, Desai U D, Pizzichini G, {\em et al.}
            Detection of a fast, intense and unusual gamma-ray 
	    transient.  Astrophys J, 1980, 237: L1 }

\end{description}


\begin{figure}[htb]
  \begin{center}
  \leavevmode
  \centerline{ 
  \epsfig{figure=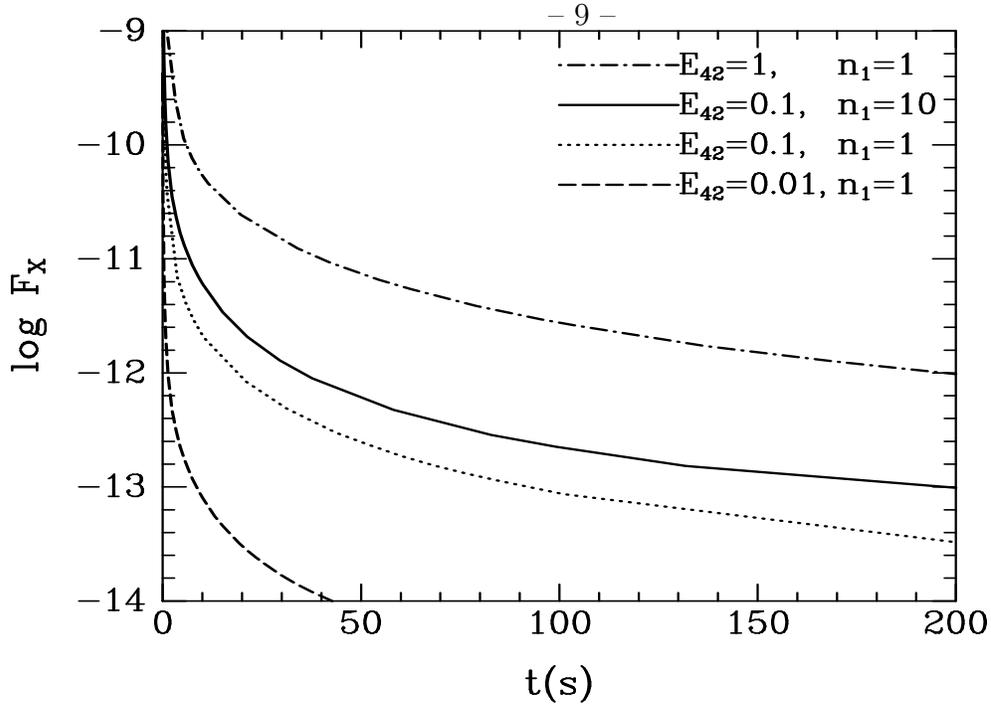,width=3.0in,height=2.5in,angle=270,
  bbllx=230pt, bblly=250pt, bburx=510pt, bbury=570pt}
  }
\caption {Predicted X-ray afterglows from SGR bursts.
Flux ($0.1 - 10$ keV) is in unit of ergs$\cdot$cm$^{-2}\cdot$s$^{-1}$  }
  \end{center}
  \end{figure}

\begin{figure}[htb]
  \begin{center}
  \leavevmode
  \centerline{ 
  \epsfig{figure=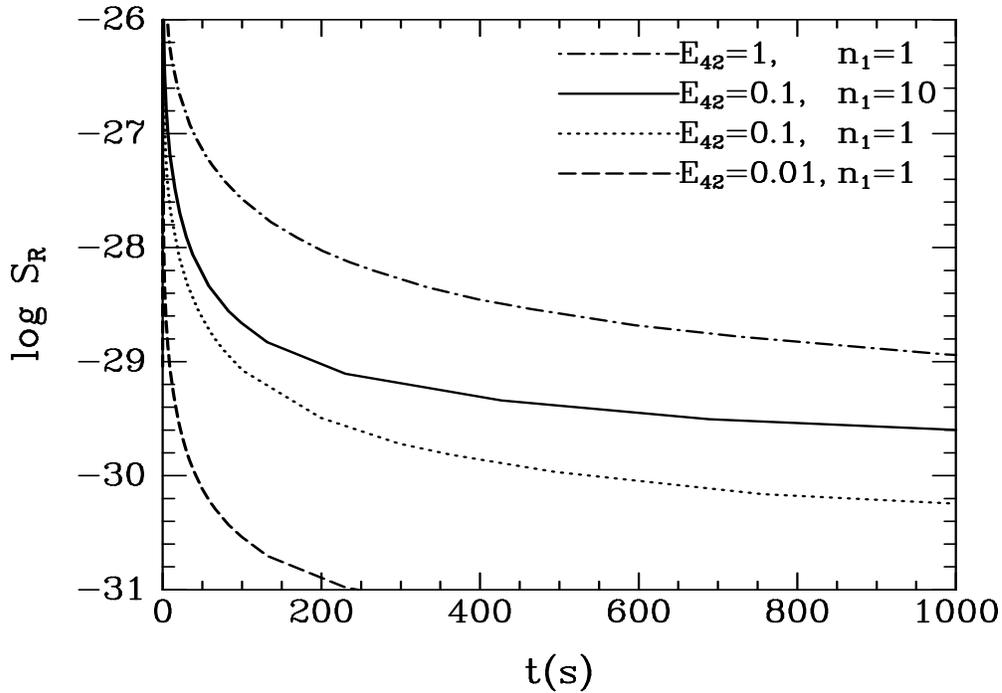,width=3.0in,height=2.5in,angle=270,
  bbllx=230pt, bblly=250pt, bburx=510pt, bbury=570pt}
  }
\caption {Predicted optical afterglows from SGR bursts. 
$S_{\rm R}$ is in unit of
ergs$\cdot$cm$^{-2}\cdot$s$^{-1}\cdot$Hz$^{-1}$  }
  \end{center}
  \end{figure}

\begin{figure}[htb]
  \begin{center}
  \leavevmode
  \centerline{ 
  \epsfig{figure=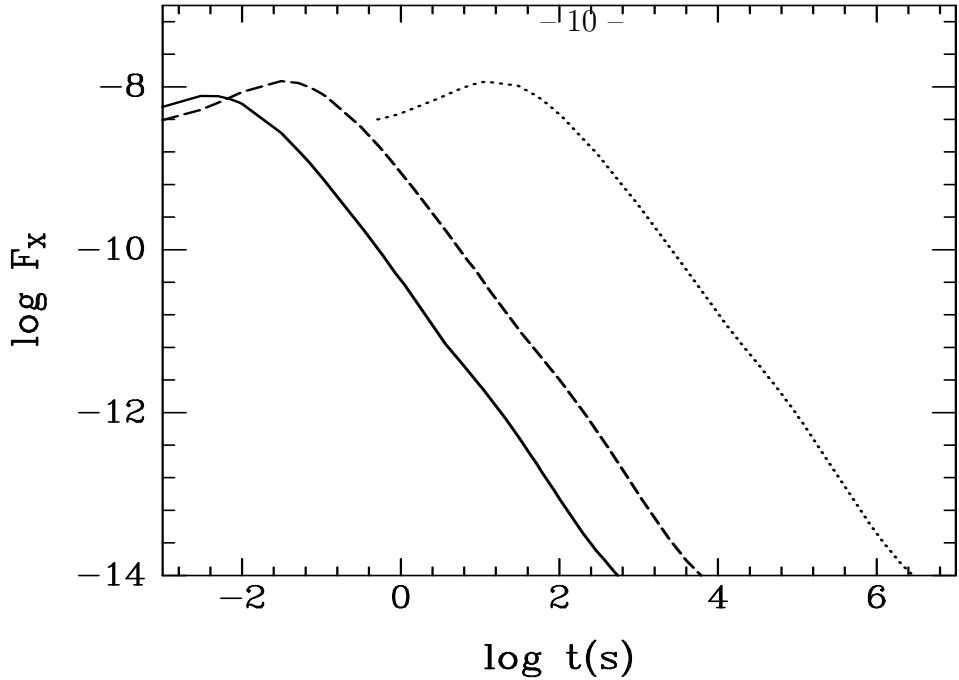,width=3.0in,height=2.5in,angle=270,
  bbllx=230pt, bblly=250pt, bburx=510pt, bbury=570pt}
  }
\caption {Theoretical X-ray afterglows from GRBs at different 
distances. The three lines correspond to cosmological 
($E_0 = 10^{52}$ ergs, $d = 3$ Gpc, dotted line),
the Galactic Halo ($10^{44}$ ergs, $300$ kpc, dashed line),
and SGR ($10^{41}$ ergs, $10$ kpc, full line) bursts 
respectively  }
  \end{center}
  \end{figure}

\begin{figure}[htb]
  \begin{center}
  \leavevmode
  \centerline{ 
  \epsfig{figure=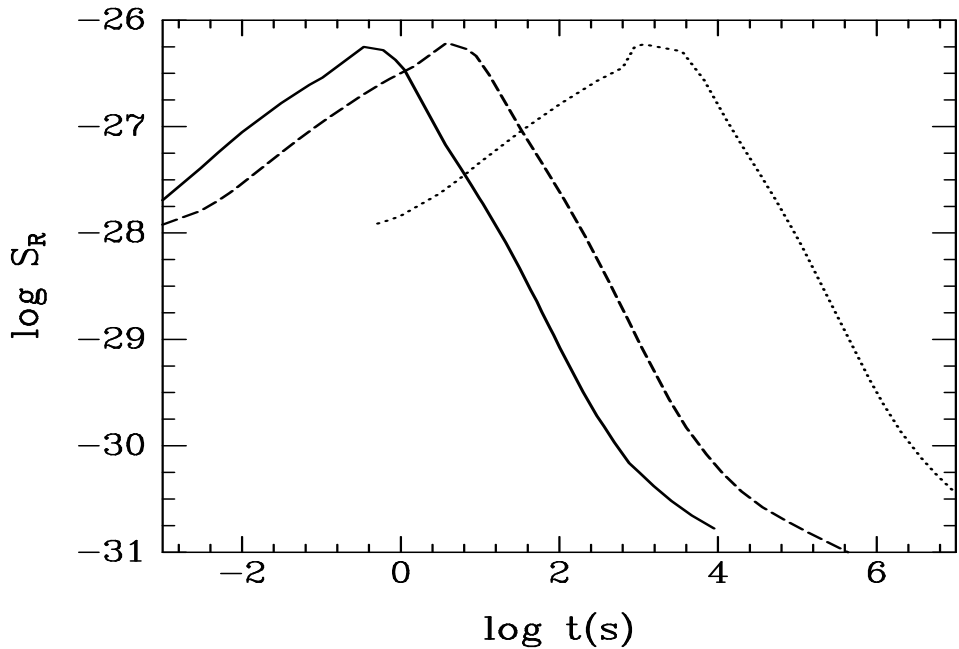,width=3.0in,height=2.5in,angle=270,
  bbllx=210pt, bblly=250pt, bburx=510pt, bbury=570pt}
  }
\caption { Calculated optical afterglows from cosmological  
(dotted line), the Galactic Halo (dashed line), and SGR (full line)
bursts }
  \end{center}
  \end{figure}

\begin{figure}[htb]
  \begin{center}
  \leavevmode
  \centerline{ 
  \epsfig{figure=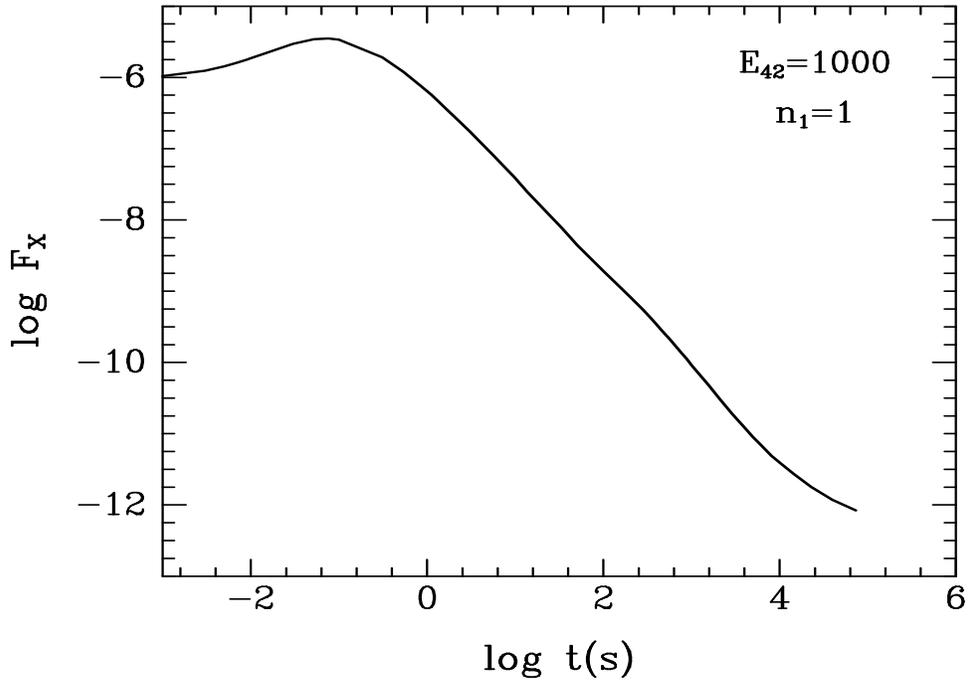,width=3.0in,height=2.5in,angle=270,
  bbllx=230pt, bblly=250pt, bburx=510pt, bbury=570pt}
  }
\caption { Theoretical X-ray afterglows from GRB 790305 }
  \end{center}
  \end{figure}
\vspace{10mm}

\begin{figure}[htb]
  \begin{center}
  \leavevmode
  \centerline{ 
  \epsfig{figure=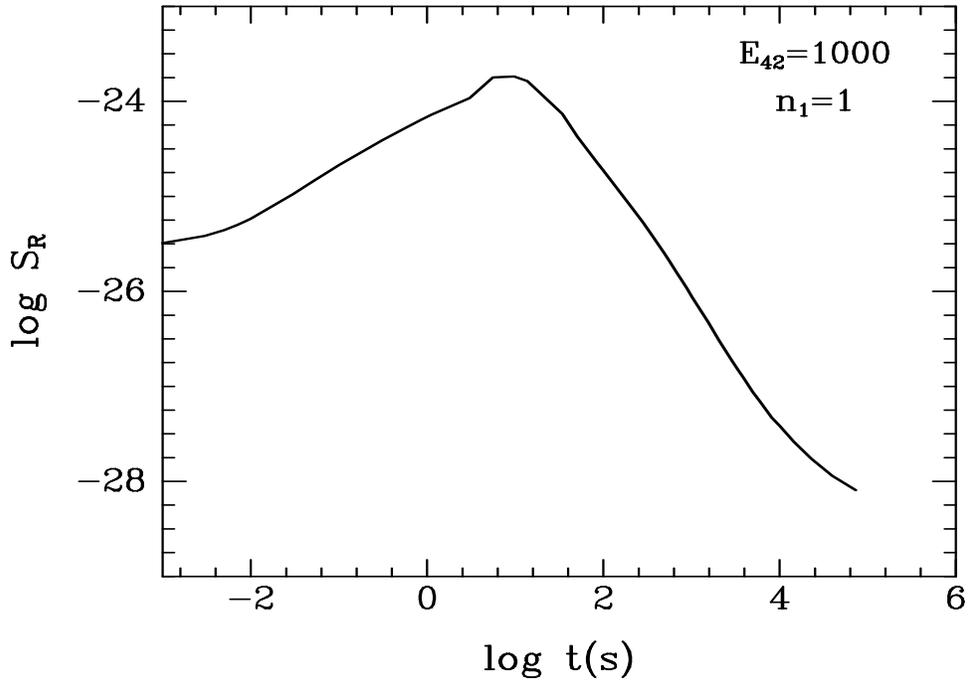,width=3.0in,height=2.5in,angle=270,
  bbllx=230pt, bblly=250pt, bburx=510pt, bbury=570pt}
  }
\caption { Calculated optical afterglows from GRB 790305}
  \end{center}
  \end{figure}
\end{document}